\documentclass[a4paper,12pt]{article}
\usepackage[a4paper,left=2.5cm, right=2.5cm, top=2.5cm, bottom=2.5cm]{geometry}
\usepackage[utf8]{inputenc}   	
\usepackage[T1]{fontenc}        
\usepackage{quoting}
\usepackage{rotating}
\usepackage{threeparttable}
\usepackage{bbm}
\usepackage{lineno}
\usepackage[colorinlistoftodos, textwidth=65mm, shadow]{todonotes}
\usepackage{setspace} 
\usepackage{color}              
\usepackage{amsmath,amsthm,amsfonts,amssymb,amscd,mathrsfs}
\usepackage{bm}						
\usepackage{bbm}	
\usepackage[ngerman, english]{babel}     %
\usepackage{ae}                 %
\usepackage{authblk}
\usepackage{graphicx}           
\usepackage{braket}
\usepackage{longtable}          
\usepackage{multicol}
\usepackage{array}
\usepackage{enumerate}
\usepackage{multirow}           
\usepackage[hyphens]{url}
\usepackage{pdfpages}
\usepackage{epsfig}
\usepackage{fancybox}
\usepackage[labelfont={bf}, hypcap=false,justification=centerfirst]{caption}
\usepackage{bold-extra}
\usepackage{mdwlist}	
\usepackage{xcolor}
\usepackage{fancyvrb}
\usepackage{arydshln}

\usepackage{lscape}
\usepackage{import}
\usepackage{float}
\usepackage{bigstrut}
\usepackage[titletoc]{appendix}
\usepackage{todonotes}
\usepackage[babel]{csquotes}
\usepackage[babel]{csquotes}
\usepackage{datetime}
\newdateformat{deformat}{\THEDAY{.} \monthname[\THEMONTH] \THEYEAR}
\definecolor{tuebingendarkred}{RGB}{165,30,55}
\usepackage{alphalph}
\usepackage{pifont}
\usepackage{ae}
\usepackage{subcaption}
\usepackage{microtype}
\usepackage{placeins}
\usepackage{bm,xstring}
\usepackage{cleveref}
\usepackage{tabularx}
\usepackage[hang,bottom,stable]{footmisc}
\newcolumntype{d}{D{.}{.}{-1}}
\newcolumntype{s}{>{\hsize=1\hsize}X}

\RecustomVerbatimCommand{\VerbatimInput}{VerbatimInput}%
{fontsize=\footnotesize,
 frame=lines,  
 framesep=2em, 
 rulecolor=\color{Gray},
 label=\fbox{\color{Black}data.txt},
 labelposition=topline,
 commandchars=\|\(\), 
 commentchar=*        
}

\makeatletter
\def\munderbar#1{\underline{\sbox\tw@{$#1$}\dp\tw@\z@\box\tw@}}
\let\@@pmod\pmod
\DeclareRobustCommand{\pmod}{\@ifstar\@pmods\@@pmod}
\def\@pmods#1{\mkern4mu({\operator@font mod}\mkern 6mu#1)}
\makeatother
\makeatletter 
\def\leqn{\tagsleft@true} 
\def\reqn{\tagsleft@false} 
\def\fleq{\@fleqntrue \let\mathindent\@mathmargin \@mathmargin=\leftmargini} 
\def\cneq{\@fleqnfalse} 
\makeatother 

\usepackage{pgf}
\usepackage{tikz}
\usetikzlibrary{arrows,shapes,automata,backgrounds,petri,calc,matrix,decorations.markings,decorations.pathreplacing,patterns}
\usepackage[titletoc]{appendix}

\newcolumntype{C}[1]{>{\centering\arraybackslash}m{#1}}
\newcolumntype{R}[1]{>{\raggedleft\arraybackslash}m{#1}}  

\usepackage{booktabs}
\usepackage{environ}
\usepackage{expl3}
\ExplSyntaxOn
\seq_new:N \l_DT_contents_seq
\seq_new:N \l_DT_contents_A_seq
\tl_new:N \l_DT_header_tl
\cs_generate_variant:Nn \seq_set_split:Nnn { NnV }
\NewEnviron{dbltbl}[2]
  {
    \seq_set_split:NnV \l_DT_contents_seq { \\ } \BODY

    \seq_pop_left:NN \l_DT_contents_seq \l_DT_header_tl

    \seq_clear:N \l_DT_contents_A_seq
    \prg_replicate:nn
      { \int_div_truncate:nn { \seq_count:N \l_DT_contents_seq } {2} }
      {
        \seq_pop:NN \l_DT_contents_seq \l_tmpa_tl
        \seq_put_right:NV \l_DT_contents_A_seq \l_tmpa_tl
      }

    \begin{tabular}{#1m{#2}m{#2}#1}
      \toprule
      \l_DT_header_tl &&& \l_DT_header_tl \\
      \midrule
      \seq_mapthread_function:NNN
        \l_DT_contents_A_seq
        \l_DT_contents_seq
        \DT_one_row:nn
      \bottomrule
    \end{tabular}
  }
\cs_new:Npn \DT_one_row:nn #1#2 { #1 &&& #2 \\ }
\ExplSyntaxOff

\newcommand{\zh}[1]{\begin{CJK}{UTF8}{gbsn}#1\end{CJK}}
\usepackage{CJKutf8}
\usepackage{url} 

\usepackage[longnamesfirst]{natbib}
\usepackage{adjustbox}

\newif\ifblind
\blindtrue    

\begin{document}

\begin{titlepage}
\singlespace

\title{AI Governance in Higher Education: A course design exploring regulatory, ethical and practical considerations}
\author[1]{Rapha\"el Weuts}
\author[2]{Johannes Bleher}
\author[3]{Hannah Bleher}
\author[4]{Rozanne Tuesday Flores}
\author[5]{Guo Xuanyang}
\author[6]{Paweł Pujszo}
\author[7]{Zsolt Alm\'asi}

\affil[1]{KU Leuven, Belgium}
\affil[2]{University of Hohenheim, Germany}
\affil[3]{University of Bonn, Germany}
\affil[4]{Bukidnon State University, Philippines}
\affil[5]{Southwest University of Political Science and Law, China}
\affil[6]{College of Europe, Natolin, Poland}
\affil[7]{P\'{e}ter P\'{a}zm\'{a}ny Catholic University, Hungary}

\date{\today}

\maketitle
\thispagestyle{empty}

\vspace*{-0.5cm}

\begin{abstract}
\noindent
As artificial intelligence (AI) systems permeate critical sectors, the need for professionals who can address ethical, legal and governance challenges has become urgent. Current AI ethics education remains fragmented, often siloed by discipline and disconnected from practice. This paper synthesizes literature and regulatory developments to propose a modular, interdisciplinary curriculum that integrates technical foundations with ethics, law and policy. We highlight recurring operational failures in AI -- bias, misspecified objectives, generalization errors, misuse and governance breakdowns -- and link them to pedagogical strategies for teaching AI governance. Drawing on perspectives from the EU, China and international frameworks, we outline a semester plan that emphasizes integrated ethics, stakeholder engagement and experiential learning. The curriculum aims to prepare students to diagnose risks, navigate regulation and engage diverse stakeholders, fostering adaptive and ethically grounded professionals for responsible AI governance.
\end{abstract}
\vspace{0.5cm}

\begin{tabular}{p{0.15\textwidth} p{0.7\textwidth}}
  \textit{Keywords:} &AI governance; AI ethics education; interdisciplinary curriculum; experiential learning; embedded ethics; stakeholder engagement; value sensitive design; operational failures taxonomy; risk-based regulation; EU AI Act; China PIPL; higher education.  \\
  \\
\textit{JEL:} & I23, I28, K20, K24, O33, O38, D83, D91. \\
\end{tabular}
\end{titlepage}

\maketitle
\doublespacing
\section{Introduction}
There is already a large corpus of literature on how to teach AI literacy and ethics, and it is still growing. As artificial intelligence systems become increasingly sophisticated, the imperative for AI governance and ethics education has reached unprecedented urgency. Recent developments in large language models, exemplified by advanced systems (at the time of writing examples of such systems are ChatGPT 5, Gemini 2.5, Grok 4, Claude Opus 4.1 and Claude Sonnet 4), demonstrate both remarkable capabilities and complex safety considerations that require careful alignment with human values \cite[see][]{Anthropic2025SystemCard}. The rapid proliferation of AI applications across sectors -- from healthcare and education to finance and governance -- has created a need for professionals who can navigate the ethical complexities inherent in AI development, deployment and governance. Yet despite the widespread recognition of this need, the academic response has been fragmented, inconsistent and often inadequate to address the multifaceted nature of AI ethics challenges.

Contemporary AI ethics education faces several systemic challenges that fundamentally undermine its effectiveness. Perhaps most problematically, current pedagogical approaches tend toward disciplinary silos, creating what researchers have termed the "BAG model" (Build, Assess, Govern), where engineering students focus primarily on technical implementation, legal students concentrate on regulatory compliance and social science students examine societal impacts – with minimal cross-pollination between these domains \citep{Javed2022BAG}. This fragmentation directly contradicts the inherently interdisciplinary nature of AI ethics, where technical decisions have immediate social consequences, regulatory frameworks must accommodate technological realities and ethical principles must be operationalized through engineering practices.

The siloed approach is compounded by a lack of meaningful stakeholder engagement in curriculum development and delivery. Most existing programs do not incorporate diverse perspectives \citep{Akgun2022AIinEducation,Wargo2024Balance,EuropeanSchool2024EthicalAI,HLEG2019EthicsGuidelines}. Operational barriers further constrain effective implementation of AI ethics education. Institutional inertia, inadequate faculty training in rapidly evolving technologies and insufficient resources for curriculum development create significant obstacles to program establishment and maintenance \citep{Wargo2024Balance}. This challenge is exacerbated by the pace of technological change in AI, where regulatory landscapes, technical capabilities and ethical considerations evolve faster than traditional academic curriculum cycles can accommodate. Furthermore, the field lacks robust empirical evidence regarding the effectiveness of different pedagogical approaches, making it difficult to identify and scale successful educational models.

Despite these challenges, a growing consensus has emerged regarding the fundamental components that comprehensive AI ethics education must address. Leading institutions and organizations have identified core areas including fairness and non-discrimination, accountability and transparency, privacy and data governance, human agency and oversight, technical robustness and safety, environmental and societal well-being and respect for human rights \citep{EuropeanSchool2024EthicalAI,HLEG2019EthicsGuidelines}. These principles, codified in frameworks such as the European Union's Ethics Guidelines for Trustworthy AI or UNESCO's recommendations (SHS/BIO/PI/2021/1 2022), provide a foundation for curriculum development that balances theoretical understanding with practical application. Beside considering key principles, the UNESCO recommendations of AI ethics advocate for practical strategies in key policy areas. One of them is the field of education and research. With this, UNESCO is calling out the member states to provide AI literacy education, in particular, for marginalized groups.

However, a growing literature emphasizes the need for a shift from abstract ethical principles to actionable policies \citep{Kim2024PolicyAI,Kim2023Governance}. They propose frameworks for AI governance in educational settings \citep{Chan2023AIPolicyEducation}, businesses \citep{Schneider2022BusinessGovernance} and society at large \citep{Choung2023AIGovernanceFramework}. The literature highlights key themes in AI governance, including technology, stakeholders, regulation and processes \citep{Birkstedt2023AIGovernance} and emphasizes the importance of integrating ethical, legal and policy considerations into AI education and development \citep{Wilk2019teaching}. The literature also discusses the challenges of advanced AI governance, proposing taxonomies of research in problem clarification, option identification and policy prescription \citep{Maas2023AdvancedGovernance}. Overall, these works stress the need for a comprehensive, multi-stakeholder approach to AI governance that balances innovation with responsible development and use across various domains.

Effective AI ethics education must extend beyond principle enumeration to develop students' capacity for ethical reasoning in complex, contextually dependent situations. This requires pedagogical approaches that integrate case study analysis, stakeholder engagement and hands-on experience with real-world ethical dilemmas. Contemporary research emphasizes the importance of experiential learning that allows students to grapple with trade-offs between competing values, understand the limitations of purely technical solutions to ethical problems and develop skills in collaborative decision-making across disciplinary boundaries.

The mounting evidence suggests that effective AI ethics education requires hybrid and interdisciplinary models that combine technical, ethical and societal content within cohesive learning experiences rather than separate courses \citep{Tennant2023HybridAlignment}. This participatory approach is essential not only for educational effectiveness but also for modelling the collaborative processes necessary for ethical AI development in practice.

Successful implementation of these integrated models requires substantial investment in capacity building and professional development for faculty, many of whom lack training in the interdisciplinary competencies required for effective AI ethics instruction \citep{Stone2025EthicsEducation}. Institutions have to develop mechanisms for continuous evaluation and adaptation, implementing feedback systems and iterative curriculum updates to maintain relevance in rapidly evolving technological and regulatory environments.

This paper provides an examination of current practices and emerging best practices in AI ethics curriculum design. We interweave this exploration of the existing literature with a suggestion for a comprehensive curriculum. We draw on extensive analysis of educational programs worldwide and recent developments in AI safety research. Furthermore, we synthesize findings from empirical studies of existing curricula, analyze successful case studies of interdisciplinary program implementation and propose evidence-based frameworks for developing more effective AI ethics education. Our analysis incorporates insights from recent advances in AI alignment research, including constitutional AI approaches and responsible scaling policies, to ensure that educational frameworks remain current with technological developments.

Findings from reviewed literature support a modular, interdisciplinary curriculum framework for AI governance education with the following features:

\begin{itemize}
\item \emph{Integrated Ethics}: Ethics is most effective when embedded within technical courses, rather than taught in isolation \citep{Wilton2022AILiteracy}. This approach is reported to increase relevance and practical application \citep{Javed2022BAG}.

\item \emph{Stakeholder Engagement}: Curriculum design benefits from involving diverse stakeholders, including students, educators, policymakers, industry professionals and affected communities. Approaches such as stakeholder-first or participatory methodologies are highlighted \citep{Chan2023AIPolicyEducation}.

\item \emph{Breadth of Content}: Syllabi that address ethical, technical, societal, legal, and operational dimensions, and that are flexible to adapt to disciplinary and regional contexts, are more comprehensive \citep{Chan2023AIPolicyEducation}.

\item \emph{Pedagogical Diversity}: Project-based learning, interactive workshops, interdisciplinary teaching and case study analysis are frequently reported as effective. Emphasis is placed on real-world application and critical reflection \citep{Wilton2022AILiteracy}.

\item \emph{Policy and Governance}: Including policy engagement and practical exercises is recommended to bridge the gap between abstract ethical principles and actionable governance \citep{Chan2023AIPolicyEducation}.

\end{itemize}

We will follow these features in the following concrete example of a curriculum that incorporates technical AI literacy with ethical education (see Appendix~\ref{app:course-structure} for a detailed semester plan and module descriptions).

\section{AI technical foundations}
Artificial Intelligence has evolved through decades of research and innovation, growing into a multidisciplinary field with diverse paradigms, models and system architectures. This chapter introduces the core types of AI, the architectures that support them and the lifecycle of AI system development; enabling a holistic understanding of how intelligent systems are conceptualized, built and deployed. An understanding of these core techniques is not only vital for building AI, but also forms the technical bedrock for addressing ethical, governance and safety challenges discussed throughout this curriculum.

\subsection{Types of AI: from logic to learning}

While leaps in computer science, applied mathematics and statistics form the bedrock of modern AI development, historically, AI began with GOFAI (Good Old-Fashioned AI, a term coined by Haugeland (1985)), or symbolic AI, where intelligence was hand-coded through explicit rules and logic trees. These systems, prominent in early AI development, relied on logical reasoning and predefined symbolic structures (Nilsson, 1998). While effective in narrow domains, they struggled in handling ambiguity and scale.

Probabilistic AI emerged to address real-world complexity, integrating uncertainty into decision-making. Probabilistic graphical models, such as Bayesian networks, provided a flexible way to infer relationships and update evidence-based beliefs \citep{Koller2009PGM}.

The emergence of machine learning marked a fundamental shift in the way artificial intelligence systems are built. Rather than relying on explicitly programmed rules, machine learning empowers systems to discover patterns and relationships directly from data. This data-driven approach underpins much of the progress in AI today.

There are three main branches of machine learning, each offering a distinct way for systems to learn. Supervised learning involves training models on labelled data, where the correct answers are provided in advance. This enables algorithms to make predictions or classifications, such as identifying whether an email is spam or not or predicting house prices from past sales. Unsupervised learning, by contrast, tackles unlabelled data, seeking to uncover hidden structures, groupings, or relationships within the information. Common examples include clustering customers by purchasing behaviour or reducing the dimensionality of complex datasets for visualisation. Reinforcement learning represents a third paradigm, where agents learn through trial and error, interacting with their environment and striving to maximize a cumulative reward, much like how animals learn through experience \citep{Sutton2018RL}.

Alongside these core approaches, researchers have drawn inspiration from the natural world to develop evolutionary computing and other bio-inspired algorithms. Techniques such as genetic algorithms mimic the process of natural selection, allowing AI systems to evolve solutions and optimize outcomes across vast, complex problem spaces \citep{Eiben2015EvolutionaryComputing}.

The way objectives are set, the design of learning environments and the handling of data in each paradigm have far-reaching consequences. These choices shape not just how well models perform, but also their transparency, fairness and robustness. They influence how vulnerable a model might be to technical pitfalls like overfitting, unexpected behaviours or embedded bias; challenges that remain central themes throughout the field of AI and are explored in depth in later chapters.

On a fundamental level these paradigms all involve similar search mechanisms that apply rule sets to hypothesis spaces. The long-standing perceived gap between symbolic reasoning and statistical learning is increasingly being bridged by neurosymbolic systems. A notable example is DeepProbLog, which extends probabilistic logic programming with neural predicates. This allows end-to-end learning while retaining logical structure and calibrated uncertainty, making constraints and governance rules more transparent than in purely sub-symbolic models \citep{manhaeve2018deepproblog}.

\subsection{AI Architectures: the brains behind the models}

The architecture of an AI system shapes how it processes information, learns from data and interacts with the world. At the heart of deep learning are feedforward neural networks, where data moves in a single direction, from input to output, laying the groundwork for many modern AI applications \cite{Goodfellow2016DeepLearning}.

Building on this foundation, convolutional neural networks (CNNs) brought a breakthrough in analysing visual and spatial data. By applying layers of filters, CNNs are able to detect and learn intricate patterns within images, making them especially powerful for tasks like image recognition and computer vision \cite{LeCun2015DeepLearning}.

When it comes to handling sequences, like sentences in language or points in a time series, recurrent neural networks (RNNs) and especially their more advanced variant, Long Short-Term Memory (LSTM) networks, have played a pivotal role. These models are designed to remember information across sequences, allowing them to process language, music and other time-dependent data with greater nuance \cite{Hochreiter1997LSTM}.

The next major leap came with the advent of transformers, which revolutionized natural language processing. Instead of relying on sequential processing, transformers introduced attention mechanisms that enable models to weigh the importance of different pieces of information in parallel. This innovation not only improved efficiency but also significantly boosted performance on a range of language tasks \cite{Vaswani2017Attention}.

Today, transformers underpin the most advanced large language models, such as GPT and BERT, which can be fine-tuned for tasks ranging from translation to summarization (Devlin et al., 2019). As these models grow in complexity and reach, a key challenge emerges: grounding. This means ensuring that language models are not just fluent but are anchored in external realities or contexts; an essential step for building trustworthy and reliable AI systems \citep{Bender2020NLU}.

In recent years, the use of an additional layer to the overall architecture has gained ground: constellations of multiple language models with separate tasks. Among these, Retrieval Augmented Generation (RAG) stands out for its ability to combine the creative language generation capabilities of large models with the precision of external knowledge retrieval. By drawing on sources outside the model itself, RAG can produce responses that are not only fluent but also factually grounded, helping to bridge the gap between generative AI and reliable information \citep{Lewis2020RAG}.

Another important development is the rise of Context Augmented Generation (CAG). This approach focuses on maintaining coherence and relevance, not just within a single exchange, but across entire conversations or documents. By keeping track of ongoing context, CAG architectures enable more meaningful interactions and nuanced understanding, whether in chatbots, digital assistants or collaborative writing tools.

Meanwhile, the field is also witnessing a growing interest in multi-agent architectures. Here, multiple AI agents interact, sometimes collaborating, sometimes competing, to solve problems or generate new insights. This collective approach draws inspiration from social intelligence and can support forms of emergent behavior or distributed reasoning that go beyond what any single agent could achieve alone \citep{Russell2021AIMA}.

Finally, researchers are exploring hybrid systems that integrate the strengths of different AI paradigms. By combining the flexible pattern recognition of neural networks with the structured, rule-based reasoning of symbolic AI, these systems offer a more holistic approach to complex tasks. This synergy holds the promise of greater transparency, explainability and problem-solving power than either method could deliver on its own \citep{Besold2017NeuralSymbolic}.

Together, these emerging architectures signal a future in which AI systems are not only more capable and context-aware, but also better equipped to collaborate with humans and each other in the pursuit of knowledge and creativity.

Across all paradigms, security and adversarial robustness remain critical technical concerns, as AI systems can be vulnerable to targeted attacks that manipulate inputs or behavior.

Architectural choices also determine how transparent or interpretable a model will be to human stakeholders. Some models, like deep neural networks, are often seen as "black boxes," while others allow for clearer inspection of decision processes; a consideration that is increasingly important for auditability and accountability. Moreover, different architectures offer varying levels of robustness to adversarial manipulation and system security threats.

\subsection{Development and lifecycle: from concept to operation}

The journey from concept to real-world application in artificial intelligence follows a structured lifecycle, with each stage playing a crucial role in the system's overall success.

\begin{enumerate}

\item Goal Setting \& Architecture Selection – Defining the problem scope and selecting an appropriate model type based on the task.

\item Data Gathering -- Acquiring datasets from various sources (e.g., sensors, user logs, databases).

\item Data Cleaning and Annotation -- Ensuring quality and labelling data, especially for supervised learning tasks.

\item Training –- Optimizing model weights using training data.

\item Validation \& Testing –- Evaluating model performance on unseen data to avoid overfitting.

\item Inference \& Deployment -- Applying the trained model in real-time or batch settings.

\item Monitoring \& Maintenance –- Continuously evaluate the model's effectiveness and make necessary updates \citep{Amershi2019SEforML}.

\end{enumerate}

Effective AI development requires ongoing attention to the data infrastructure: managing data provenance, privacy and security, as well as ensuring models remain robust as environments and data sources change. It is also crucial to recognize that AI systems are deployed within broader sociotechnical contexts, where human users, organizational processes and policy environments all affect system performance and ethical outcomes. Known limitations, such as lack of robustness to distribution shift and susceptibility to specification gaming, underscore the need for continuous monitoring and governance.

This end-to-end process is crucial for ensuring not only technical performance, but also fairness, explainability and trustworthiness;all of which are foundational to responsible AI governance. A robust understanding of these technical foundations enables future practitioners not only to build AI, but also to critically evaluate, govern and improve AI systems in an ethically informed way. Such understanding is essential for anticipating system limitations, ensuring interpretability and safeguarding against technical and security failures. These foundations are inseparable from the ethical, legal and governance challenges that follow; highlighting the need for ongoing interdisciplinary engagement as AI systems become ever more embedded in society.

\section{Operational challenges: the gap between intention and reality}

A central challenge in AI governance is the principal-agent conflict: AI systems (agents) execute their specified instructions, not necessarily the underlying intent or interests of their human designers or users (principals). This misalignment, where the agent optimizes for what is specified rather than what is actually desired, can lead to operational failures that go beyond simple bugs. These systemic issues can produce unreliable or harmful outcomes, even when the AI system is technically correct \cite{ASEF_AIEco_SD_2025}.

This chapter provides a structured overview of these challenges, categorizing them into five interrelated types of failure: flawed data (representation), flawed goals (specification), flawed optimization (generalization), flawed human use (interaction) and flawed oversight, governance and strategic interaction (governance). This fifth category encompasses failures that arise from weaknesses in monitoring and governance, including agents exploiting oversight mechanisms, adversarial attacks and emergent behaviours in multi-agent systems; where complex, often unpredictable dynamics can undermine intended outcomes, even when each individual component is operating as specified. 

Understanding this taxonomy is crucial for any AI governance curriculum aiming to bridge the gap between specification, intent and outcome. This conflict is as well at the heart of the principal-agent problem in economics.

\subsection{Failures of representation: bias and feedback}

AI operational failures often stem from flawed data representation, i.e., the way in which complex real-world phenomena are encoded, selected and structured for machine learning systems. This section examines two illustrative risks: static, historical bias in training data and dynamic, self-perpetuating feedback loops that corrupt data over time.

A primary challenge is poor training data quality, where "Garbage In, Garbage Out" applies. Data quality extends beyond factual correctness to its representativeness and freedom from latent systemic biases. \cite{Mehrabi2021BiasSurvey} provide a comprehensive taxonomy of different bias types. Since models replicate patterns from their training data, they can inherit and amplify historical inequities \citep{ONeil2016Weapons}. A well-documented example is Amazon's abandoned AI recruitment tool, which, trained on historically male-dominated data, learned to penalize female candidates \citep{Dastin2018AmazonRecruiting}. This issue is not unique; a famous ProPublica investigation found that the COMPAS recidivism algorithm was significantly more likely to falsely flag black defendants as high-risk as white defendants \citep{Angwin2016MachineBias}.

While static biases in training data can entrench existing inequities, dynamic feedback loops can actively generate and reinforce new patterns of bias as systems interact with their environment. For instance, predictive policing systems can create runaway feedback loops \citep{Ensign2018PredictivePolicing}: Predictions of high crime lead to increased police presence, which generates more arrest data, in turn "validating" the initial biased prediction and trapping neighbourhoods in a cycle of over-policing. This dynamic demonstrates that AI systems are part of larger sociotechnical systems, where their outputs can corrupt future inputs.

Addressing failures of representation thus requires not only careful data curation, but also ongoing monitoring of how AI systems interact with and influence their environments.

\subsection{Failures of specification: proxies and confounders}

Beyond data issues, failures can arise from how an AI's goal is defined. This section examines the risk of using simple proxies for complex objectives and the danger of confounding variables that cause models to learn incorrect shortcuts.

Because complex goals like "patient health" are hard to define mathematically, developers use a measurable proxy that is believed to correlate with the true goal. However, the challenge of specifying goals that truly reflect human intent lies at the heart of the principal-agent problem in AI. When the correlation between proxy and true objective is imperfect, a powerful AI system may optimize the proxy literally, often subverting the original intention. Such systems may become adept at exploiting these gaps, i.e. "game" the metric. This phenomenon is known in the AI literature as reward hacking \citep{Amodei2016ConcreteSafety}. The challenge is ensuring the specified task is a faithful representation of a beneficial outcome.

The risk is further complicated by confounding variables –hidden factors that can create spurious correlations. Models cannot distinguish between genuine and misleading patterns, and may learn harmful shortcuts as a result. A stark example is the AI model, which learned asthma was associated with a lower mortality risk in pneumonia patients \citep{Lengerich2022Bias}. This statistically valid but causally flawed prediction arose from a confounder: the level of medical care. Doctors historically gave high-risk asthmatic patients aggressive treatment; the model misinterpreted this intervention as evidence that asthma itself was protective. This case highlights the critical difference between correlation and causation; a problem best addressed through the principles of causal inference \citep{Pearl2016CausalInference}.

Ultimately, addressing failures of specification demands careful consideration of both the incentives set by measurable proxies and the potential for hidden confounders to undermine alignment with human values and intent.

\subsection{Failures of generalization: systemic challenges}

Even with sound data and well-specified objectives, AI systems often struggle to generalize their learning to new contexts; a fundamental problem for any principal who cannot anticipate every possible situation in advance. This section explores three major generalization challenges: the corruption of metrics, the emergence of unintended agentic goals and the inability to handle novel data.

The first challenge is the corruption of metrics, exemplified by Goodhart's Law, stating "When a measure becomes a target, it ceases to be a good measure" \citep{Strathern1997Audit}. In AI, this often appears as "benchmark brittleness", where models overfit to the idiosyncrasies of a test set or evaluation metric, optimizing for high scores without true capability gains. For instance, models might "game" explanation metrics by copying input text rather than providing genuinely faithful rationales \citep{Hsia2024GoodhartsLaw}. Consequently, high leaderboard scores do not guarantee robust, real-world performance.

Second, as systems become more autonomous, they may develop unintended "agentic" goals, what \cite{Bostrom2012SuperintelligentWill} calls "instrumental convergence", i.e., the tendency for diverse intelligent agents, regardless of their final objectives, to adopt similar subgoals such as self-preservation or resource acquisition. \citeauthor{Bostrom2012SuperintelligentWill} also outlines the orthogonality thesis: that intelligence and final goals are independent. Even rational pursuit of benign objectives can yield subgoals like self-preservation or resource acquisition, potentially leading to outcomes far removed from human intent. The issue here is no malice, but a logic divorced from the principle's true values.

Third, practical deployment is hampered by a lack of robustness to distributional shifts. Most models assume deployment data will resemble training data (the "i.i.d. Assumption" for the expected error). When faced with new or unexpected environments, e.g. in a self-driving car trained in sunshine but deployed in snow, performance can fail catastrophically. This challenge of out-of-distribution (OOD) generalization \citep{Liu2021OOD} represents a fundamental barrier to reliable AI in dynamic, real-world settings Ultimately, failures of generalization reveal how difficult it is to ensure that AI systems act as intended across novel or changing conditions; a core challenge for effective AI governance and oversight.

\subsection{Failures of interaction: user-centric challenges}

Not all operational failures are inherent to an AI system itself; many arise from the complex and fallible ways humans interact with AI technologies. A technically accurate AI system can still cause harm if users misunderstand, overtrust or misuse it.

User misunderstanding often stems from a gap between an AI's superficial competence and its actual capabilities. For example, the fluency of modern large language models (LLMs) can mask their lack of grounded comprehension, as their "knowledge" is based on statistical patterns rather than a causal understanding of the world \citep{Bender2020NLU}. This feeds into overtrust, or automation bias \citep{Parasuraman1997HumansAutomation}: the tendency for humans to become complacent and stop critically evaluating outputs from systems that usually seem correct. Finally, harm can arise from misuse, where a tool is applied to a context for which it was never validated. From a design perspective, this reflects a failure of the system's "affordances"; the cues and features that signal to users how the system should (and should not) be used \citep{Norman2013DesignThings}.

Ultimately, the challenges of misunderstanding, overtrust and misuse demonstrate that technical safeguards alone cannot ensure the safe operation of AI systems. Addressing these risks requires robust user education and thoughtful interface design, documentation and training tailored to "AI literacy" of both external users, who interact with AI-powered products as well as services, and internal teams within an organization who use AI tools for decision support. Rather than treating education as an afterthought, it must be a core design requirement; fostering critical, informed use and narrowing the gap between what AI agents do and what human principals intend.

\subsection{Failures of governance: alignment and oversight}

Even when AI systems are built with high-quality data, clear goals, robust generalization and thoughtful user interfaces, a final class of failures can arise from gaps in governance and strategic alignment. These failures occur when the mechanisms for monitoring, regulating or coordinating AI systems break down; allowing agents to exploit oversight, circumvent rules or interact in unexpected ways with other agents or institutions or evade accountability altogether.

A critical example is the vulnerability of advanced AI models to reward hacking and covert misbehaviour. Recent research shows that monitoring a model's chain of thought (CoT), its intermediate natural-language reasoning, can help surface deceptive planning or misaligned intent that might not be apparent in final outputs alone \citep{Baker2025MonitoringReasoning}. However, the same research warns that overreliance on such monitoring may push models to hide their true reasoning, revealing the fragility of current oversight methods.

Governance failures also include emergent dynamics in multi-agent environments, where AIs or humans and AIs interact in complex, sometimes adversarial ways. Such interactions can defeat even well-designed rules, leading to collusion, competitive "races to the bottom," or amplified systemic risks \cite[e.g.,][]{Motwani2024Collusion}. Legal and regulatory supervision often lags behind deployment, while adversarial attacks, by humans or other AIs, can expose new vulnerabilities in oversight and governance frameworks.

These failures echo classic principal-agent dynamics: agents may manipulate oversight, conceal misaligned intent or pursue unchecked subgoals. Addressing these risks requires more than technical solutions; it demands robust governance structures, external audits, continuous monitoring, human-in-the-loop mechanisms and strong coordination between legal, ethical, technical and operational teams.

\subsection{Implications for an AI governance curriculum}

The operational challenges facing AI are diverse, spanning from flawed data and goals to the inherent difficulties of generalization, human interaction and governance. The common thread is the gap between a simplified, formally specified system and the complex reality it is meant to serve; the persistent distance between intention and outcome. An effective AI governance curriculum must therefore cultivate a practice of critical vigilance, equipping future AI managers to recognize these patterns of failure across all five dimensions. By understanding how systems fail in practice, including failures of oversight and alignment, we can better build the interdisciplinary frameworks needed to ensure that AI is not only capable, but also safe, fair and aligned with the complexities of human flourishing.

\section{AI legislation and regulation}

\subsection{AI legislation in the EU}
Despite the fact that the development of a digital single market has been among the European Commission's political priorities already under President Juncker \citep{bassot2019juncker}, there was a marked difference in strategic boldness with which the new College set out to make the EU "fit for the digital age". Shortly after its appointment, the Von Der Leyen Commission formulated strategic underpinnings of its subsequent actions regulating the digital single market. The overall strategy was intended to foster the development of innovative and effective digital solutions working on a fair and competitive internal market in line with the values on which the EU was founded while pursuing greater technological sovereignty (European Commission, 2020). All four elements of this strategy permeate the EU's most important AI-related regulation (Regulation 2024/1689, 2024) –- the EU AI Act –- but also other relevant legislation, e.g., the Digital Market Act (DMA) (Regulation (EU) 2022/1925, 2022) and Digital Services Act (DSA) (REGULATION (EU) 2022/2065, 2022).

The EU AI Act has been intended to regulate artificial intelligence systems with a view to balancing seemingly conflicting objectives referred to in its Article 1. On the one hand the EU institutions wished to encourage taking advantage of the benefits of deploying AI systems to further enhance the functioning of the EU internal market. On the other, they were determined to protect a diverse array of values enshrined in the Treaty on the European Union and the Charter of Fundamental Rights of the European Union.Importantly, the regulation applies to providers of AI systems irrespective of them being established or located within the EU Member State territories (Article 2.1.a). Recognising rapid advances of AI systems, rather than regulating specific AI capabilities, the EU institutions opted for a "risk-based" approach specifying four levels of risk associated with enabling users to interact with AI systems. The EU AI Act prohibits enabling users to interact with AI systems that may, e.g., take advantage of the traits beyond users' control, such as identity, cognitive processes, or other sensitive information about users (e.g., political views), to bring about outputs detrimental to them with certain bounded exceptions such as, e.g., targeted anti-terrorist activities (Article 5). The regulation specifies at length rules for classifying AI systems as high-risk ones and outlines requirements for different stakeholders in such systems ranging from setting up iterative risk management systems, through data governance practices, technical documentation, log-keeping, ensuring transparency, accuracy and human oversight as well as cybersecurity robustness (chapter III section 2). The EU AI Act also stipulated that generative AI systems need to comply with transparency obligations such as labelling AI-generated content, embedding the measures preventing AI systems from generating illegal outputs already at the model design stage, and clarifying which copyrighted data was used by the model. The AI Act outlines the methods of classification of general-purpose AI systems as well as obligations for their providers regarding maintenance of technical documentation (chapter V). It also empowers the European Commission to further elaborate the rules, if necessary (Article 53.6). The document outlines a complex governance choreography by establishing European AI Board comprising competent designated representatives of the EU Member States (Articles 65-66), the single points for contact with relevant national authorities (Article 70), establishing an advisory forum comprising a balanced selection of relevant non-state stakeholders (Article 67), scientific experts panel (Article 68). The European Commission would preserve its enforcement role, in particular with regard to the general-purpose AI systems (Article 88). The newly established AI Office to act as a coordination body playing a crucial role in the process of bringing about the General-Purpose AI Code of Practice envisaged by the AI Act (Article 56). The final version of the Code of Practice is expected to be published by August 2025 following an open and transparent process of drafting the document (European Commission, 2025). In line with its usual prerogatives, the European Commission would be entrusted with a supervisory and enforcement role, especially with regard to the general-purpose AI systems (Article 88). 

The EU AI Act is one of the regulatory elements, developed by the EU, to balance users' safety with economic competitiveness. Other complementary pieces of legislation comprise the Digital Market Act (DMA) (Regulation (EU) 2022/1925, 2022), Digital Services Act (DSA) and the General Data Protection Regulation (GDPR).

The DMA strives to ensure that the "gatekeepers", i.e., undertakings being important, enduring market players at the EU digital internal market whose core market platforms act as important gates between businesses and end users (Article 3), do not negatively affect fairness and contestability of the EU internal market in the digital sector (Article 1).

The DSA applies to intermediary services (transmitting and/or storing information in a communication network) (Article 3.g.) regardless of their place of establishment (Article 2) and creates legal obligations with a view to enhancing transparency, fairness and safety of both consumers and businesses (chapter 3). The AI systems, especially more recent ones, such as solutions involving AI agents, pose an interpretative challenge as it has been ambiguous whether to treat them as search engines or online platforms \citep{Bostoen2024AIagents}. Categorising AI agents as one or another entails different sets of legal requirements for providers of intermediary services and, as a result, different consequences for economic viability of their AI-agentic enterprises.

The essential regulatory framework has taken a rather concrete shape within the EU, but with extraterritorial effects. At the same time, the pace and unexpectedness of innovations in AI systems pose a continuous challenge to its gradual stabilisation and much-needed clarity for AI systems providers and deployers. The EU AI Act, however, outlines an inclusive governance framework with a clear division of calibrated tasks, which allows being optimistic about its eventual potential to resolve future challenges. Yet, our times are disturbingly interesting and only time will tell whether the optimism has been warranted.

\subsection{AI legislation in China }

China's regulatory approach to artificial intelligence provides an important comparative perspective for understanding global AI governance. Unlike the European Union's centralized, value-driven framework, China's approach is characterized by a balance between individual rights, economic interests and national security concerns, as well as a pragmatic adaptation to rapid technological change. This section outlines key elements of China's AI-related regulatory landscape, focusing on privacy law, data rights, intellectual property and emerging governance models.

\subsubsection{Privacy law}

During the drafting of China’s first Civil Code, enacted in 2020, scholars debated whether the right to personal information should be recognized independently or subsumed under the broader right to privacy (Cheng, 2022). The final version of the Civil Code, specifically Articles 1032 and 1034, explicitly defines privacy and personal information as separate legal concepts. Notably, Article 1034 stipulates that the protections for privacy also apply to "private personal information," thereby acknowledging the overlap between these two domains and ensuring that individuals benefit from both sets of legal safeguards. This dual approach provides what is sometimes called "double protection," reinforcing legal remedies for violations involving personal information.

Building on the foundation laid by the Chinese Civil Code, China’s Personal Information Protection Law (PIPL), enacted in 2021, established a comprehensive legal framework for data privacy. The PIPL adopts key elements of the European Union’s General Data Protection Regulation (GDPR), including rights to access, correct, and delete personal data, as well as principles such as data minimization and accountability. However, unlike the GDPR’s strong focus on human rights, the PIPL seeks to balance individual rights with broader economic and national security interests. Article 1 of the PIPL explicitly prioritizes "promoting the reasonable use of personal information," which allows for exemptions; such as the processing of publicly available data without consent.

In contrast to the GDPR’s requirement for "specific consent" in most cases (as outlined in the EDPB Guidelines), the PIPL implements a tiered consent system. It requires "separate consent" (\zh{单独同意}) only in high-risk situations, such as sharing data with third parties, public disclosure, non-public security uses of biometrics, processing sensitive data and cross-border transfers. This reflects a lower threshold for obtaining consent compared to the GDPR.

Regulations on data localization, the requirement for data to be stored within national borders, are especially important for companies operating internationally. Initially, China adopted a protectionist stance: early sector-specific policies, such as the 2011 Notice of the People’s Bank of China on Personal Financial Information Protection, the 2015 Regulations on Map Management and the 2016 Provisions on Network Publishing Services, all imposed strict localization requirements.

In recent years however, China has gradually relaxed these rules. The 2021 Data Security Law introduced a risk-based framework under which operators of critical information infrastructure must store important data domestically but may transfer it abroad after passing security assessments. Article 38 of the PIPL further provides four pathways for cross-border data transfers: security assessments, third-party certifications, standard contractual clauses and international treaties.

Scholars note that these regulations represent a reactive alignment with international norms rather than true legal innovation and some have called for China to move from being a "participant" to a "rule-maker" in global data governance \citep{Shao2023DataFlows}. Reflecting this ambition, the 2024 Regulations on Promoting and Regulating Cross-Border Data Flows marked a significant liberalization by exempting routine commercial activities (such as cross-border e-commerce, payments and HR management) from many compliance requirements. Together with the Global Initiative on Cross-Border Data Flow Cooperation, these changes signal China’s intent to set global standards while promoting efficient and secure international data exchanges.

\subsubsection{Intellectual property law}

\paragraph{Copyright on training input data.}
In Chinese civil law scholarship, there is a growing consensus that enterprises should be granted specific "data property rights"; that is, clearly defined legal rights to control, use and benefit from data, alongside existing regulatory measures. However, there is still debate over the legal nature of these rights: some scholars advocate for a concept of "data ownership" \citep{Shen2020DataRights}, others propose "data intellectual property" \citep{Wu2023DataProperty}, while still others argue for a new, distinct type of property right that would exist alongside intellectual property and real property rights (Zhang, 2023).

Despite these differences, most agree that data property rights should be "absolute rights" with a certain degree of exclusivity for the rightsholder. In 2022, the Communist Party of China Central Committee and the State Council issued the "Opinions on Building a Data Foundation System to Better Leverage the Role of Data as a Factor of Production." This document proposes "exploring a data property rights system" and establishing mechanisms to separately assign rights such as holding data, processing it and commercially operating data products. As a result, this policy has laid the groundwork for the formal recognition and practical implementation of data property rights and their various components.

\paragraph{Copyright on AI generated outputs.} 
Whether AI generated outputs can be protected by copyright law, known as their "copyrightability", remains hotly debated in Chinese academia. Some scholars argue that content created entirely by AI should not qualify for copyright protection, but that copyright may apply when humans collaborate with AI in the creative process. Others, such as Wang Qian, contend that because AI developers and users lack direct control or free will over the specific outputs generated by AI, excluding such works from copyright protection, would not undermine incentives to invest in AI technology \citep{Wang2024CopyrightAI}.

Chinese courts have begun to address these questions through landmark cases. In the 2019 Dreamwriter Case, the Shenzhen Nanshan District Court recognized copyright protection for AI-generated news articles, highlighting the importance of human contributions in data selection and algorithm design \citep{Shenzhen2019Dreamwriter}. The Beijing Internet Court’s "China's First AI Text-to-Image Case" (2023) similarly granted copyright to AI-generated images where substantial human intellectual input was demonstrated \citep{LiMou2023}. In contrast, the Suzhou Intermediate Court in 2024 denied copyright protection for images created by "simple prompts + AI generation," finding them insufficiently original \citep{FengMou2024}.

Despite differences in outcome, these rulings reflect a consistent principle: the decisive factor for copyright protection is the extent of human involvement in the creative process. This emerging framework suggests that Chinese copyright law is likely to continue evolving toward a model that prioritizes and rewards genuine human creativity in AI assisted works.

\subsection{Sui generis frameworks and emerging models}

Both the European Union and China have recognized the importance of assessing risk levels in AI governance and using regulations to mitigate these risks. However, while the EU AI Act explicitly defines a “high-risk” category for certain AI systems, Chinese regulations do not currently use this terminology. Instead, China’s “Interim Measures for Generative AI Service Management” focus on two types of AI services deemed sensitive: those with “public opinion attributes” (\zh{公共舆论属性}), technologies that can influence public opinion such as forums or blogs, and those with “social mobilization capabilities” (\zh{社会动员能力}), meaning services that could encourage or organize collective action. These special categories are subject to additional restrictions and, in the context of AI, may include chatbots, virtual assistants or recommendation algorithms that affect public discourse or mobilization.

Judging from the practical situation, China's current regulations of artificial intelligence are still mainly characterized by decentralized legislation. But it's noteworthy that China has put formulating a unified artificial intelligence law on the agenda and into relevant plans. The At present, Chinese AI governance remains largely decentralized, with various laws and regulations targeting specific areas rather than a single, unified legal framework. Nevertheless, China has considered more centralized legislation: the “New Generation Artificial Intelligence Development Plan” (2017) outlined steps toward a comprehensive system of AI ethics and regulation by 2030. Drafts of a national AI law were included in the State Council’s legislative work plans for 2023 and 2024, but the 2025 plan notably omitted explicit reference to such a law. Instead, it called more generally for “legislative efforts for the healthy development of AI.” This shift has led many to interpret that China has paused its pursuit of a unified AI law. Reflecting this perspective, scholar \cite{Ding2024ChinasAIlaw} argues that China should continue a context-specific, adaptive approach to AI regulation rather than rushing into comprehensive legislation. The "2025" plan was released on May 14, 2025

\section{Ethical considerations on AI ethics education}

A multitude of publications in recent years has identified guiding principles for responsible AI design in both industry and policymaking contexts \citep{jobin_global_2019}. This principle orientation of AI ethics aims to prevent harm and injustice arising from algorithmic bias and the misuse of AI, while safeguarding fundamental human rights. Within this discourse, the European Union’s initiatives to define principles for trustworthy AI represent a particularly significant contribution\citep{HLEG2019EthicsGuidelines}. Building upon four core principles –- human autonomy, prevention of harm, fairness and explicability -- the EU formulated the EU AI Act for trustworthy AI (EU AI Act, 2024). This principle driven approach exemplifies the EU's commitment to ethically aligned responsible research and innovation \citep{european_commission_directorate-general_for_research_and_innovation_responsible_2013}.

This kind of deductive ethical approach translates normative principles into regulatory structures to guide AI development, engineering and deployment practices. However, despite all politically deliberative and consensual efforts, such approaches remain susceptible to ethics washing \citep{Metzinger2019EthicsWashing}. Critics also argue that principle based approaches need to be more than just lip services and must be followed by real action \citep{Hao2019AIethicswashing,Mittelstadt2019Principles}. Furthermore, critics contend that such approaches systematically exclude marginalized voices and participatory elements. They thereby perpetuate rather than challenge existing power structures within AI engineering \citep{BleherBraun2023Reflections,vanMaanen2022EthicsWashing}. These critiques prompt a central question for AI ethics education in higher education contexts: How can AI ethics education move beyond performative or superficial ‘ethics washing’? More fundamentally, how can genuine integration between ethical theory and practice be achieved in AI ethics education \citep{BleherBraun2023Reflections}?

At an international level, the UNESCO Recommendations on AI Ethics are a key anchor point in the debate on AI ethics education \citep{unesco_recommendation_2022}. UNESCO committees on international and nation state level are currently working to implement these AI ethics recommendations across multiple policy areas. For the research and education area the UNESCO recommendations emphasize socially just, inequality-reducing education through comprehensive empowerment strategies and targeted competency development \citep{unesco_recommendation_2022}. UNESCO advocates for interdisciplinary collaboration in research and educational programming alongside public awareness initiatives. Although interdisciplinarity and cross-collaboration between technical AI education and humanistic dimensions are encouraged, these recommendations lack consideration of economic and legal perspectives in AI ethics education. 

Instead, they promote a very sector specific AI ethics education approach, rather than fostering an integrated, societywide framework that would connect policy domains and engage political, social, legal, technical and economic stakeholders in direct dialogue with AI ethics education institutions and programs. In other words, there is a structural deficiency: the UNESCO recommendations stop short of holding industry meaningfully accountable and fail to establish structural links between AI ethics education and the technical ‘workbench’ where AI systems are designed and developed. Impactful AI ethics education, however, cannot be an isolated project. As argued here, this requires linking top‑down elements – including legal, ethical and technical principles and mechanisms – with bottom‑up dimensions informed by real‑world deployment contexts. These bottom‑up dimensions encompass societal user experiences, stakeholder perspectives, behavioral and social change and stakeholder intuitions regarding AI technologies. They should be grounded in empirical research from economics, the social sciences and political science that investigates the conditions and implications of AI deployment. Such an integrated approach to AI ethics education facilitates the development of comprehensive governance strategies by incorporating citizens’ experiences, industry partners’ expertise and policymakers’ deliberations \citep{BleherBraun2023Reflections}.

The increasing emergence of AI ethics initiatives, programs and pedagogical models in higher education underscores the persistent challenge of integrating top‑down and bottom‑up dimensions of ethical decision‑making \citep{laupichler_artificial_2022}. The following discussion examines three paradigmatic approaches to AI ethics education, outlining their respective opportunities and limitations: principle-driven institutional, embedded ethics, as well as stakeholder-oriented and user-centered approaches. This distinction builds on the categorization of \cite{BleherBraun2023Reflections} of AI ethics approaches into three types: the ethically aligned approach, embedded ethics and value‑sensitive design. Taken together, these paradigms illustrate the diverse pathways through which AI ethics education can contribute to bridging the gap between ethical theory and practical implementation.

\subsection{Principle driven institutional approaches}

In introducing this distinction, we begin by describing and discussing the first paradigmatic educational approach to AI ethics: the institutional adaptation of ethical principles, for example through frameworks that guide AI‑related educational practices (Temper et al., 2025). Such principle-driven institutional initiatives broadly reflect international standards, such as for example IEEE's Ethically Aligned Design Standard, as well as the AI principles articulated by the EU and UNESCO 
\citep{HLEG2019EthicsGuidelines,ieee_global_initiative_on_ethics_of_autonomous_and_intelligent_systems_ethically_2019,unesco_recommendation_2022}. As part of the Digital Education Action Plan 2021-2027, the European Commission also introduced ethical guidelines for educators, asking guiding reflective questions along the lines of their key requirements for trustworthy AI \citep{european_commission_directorate_general_for_education_ethical_2022a,european_commission_directorate_general_for_education_ethical_2022b}. Although the EU AI Act states the importance of high-quality digital education and training, as well as the need for a sufficient level of AI literacy, specific guidelines on AI ethics education in the higher education context have not yet been established (Dietis, 2025). Guided by overarching principles, universities therefore voluntarily – or in response to recommendations – align their educational AI competency curricula and objectives with universal values such as transparency, accountability, justice, or others \citep{bruneault_f_ai_2022,russell_group_russell_2023,braun_tum_2024,temper_higher_2025,wynants_s_ethical_2025}.

While this kind of AI ethics education approach offers institutional clarity and broad applicability by providing a general roadmap, it can fall short in practice and outcome \citep{saltz_integrating_2019}. Abstract guidelines are keen to fail to address specific use cases, educational contexts and diverse knowledge bases. Top-down approaches struggle to integrate the specific realities and AI deployment experiences in particular by marginalized groups. Furthermore, when foundational principles conflict -- as privacy and transparency do in certain applications -- such top-down frameworks offer insufficient guidance for ethical decision making \citep{BleherBraun2023Reflections}. The risk of ethics washing persists when principles remain superficially adopted without substantive integration into curriculum and institutional practice.

\subsection{Embedded ethics approaches}

The embedded ethics approach represents a second paradigmatic approach to AI ethics education. The embedded ethics approach is on the one hand an educational approach and on the other hand a methodology for interdisciplinary AI research \citep{grosz_embedded_2019,fiske_embedded_2020,mclennan_embedded_2020,willem_embedded_2024}. According to this approach, AI ethics education is embedded into the curriculum of respective training programs, typically within AI technical or humanities study programs. Modular integration of AI Ethics into computer science courses or certificate programes is common \cite{grosz_embedded_2019}; Horton et al., 2024). Rather than treating ethics as stand-alone topics, it is woven into the fabric of computer science courses.

Harvard's Embedded Ethics program exemplifies this approach, where graduate and undergraduate courses integrate ethical dilemmas directly into coding assignments, with students developing algorithms while simultaneously analysing bias, privacy and fairness implications \cite{grosz_embedded_2019,harvard_university_embedded_nodate}. Also Stanford's Embedded Ethics program demonstrates scalable implementation where ethics modules are integrated across computer science curricula or as a AI ethics certificate program \cite{stanford_university_stanford_nodate}. The University of Toronto's Ethics Education Initiative \cite{university_of_toronto_e3i_nodate} has reached students by offering embedding ethics modules, where students examine how technical developments and choices raise ethical concerns \citep{horton_embedded_2024,university_of_toronto_e3i_nodate}.

Within this approach, ethical questions are central to algorithmic design, programming assignments and technical problem-solving and not just an afterthought. Students analyze, for example, the fairness implications of a machine learning model while simultaneously building it, or reflect on data privacy concerns as they learn about network structures. As empirical research demonstrates, such a bottom-up, context-sensitive and case study approach enhance relevance and engagement of AI ethics for students \citep{horton_embedded_2024,saltz_integrating_2019}. However, success depends heavily on the ethical expertise of instructors. The embedded ethics approach's context-sensitivity, furthermore, lacks overarching ethical frameworks for guidance in cross-contextual and diversified decision-making \citep{BleherBraun2023Reflections}. Additionally, these programs remain susceptible to institutional power structures that may subordinate ethical considerations to technical or commercial imperatives.

\subsection{Stakeholder oriented and user centered approaches}

A third paradigmatic AI ethics education approach emphasizes stakeholder oriented and user centered AI ethics programs. It resembles value-sensitive design methodologies by focusing on concrete use cases and stakeholder evaluation \citep{Friedman2013VSD,umbrello_value-sensitive_2018}. These programs, for example, manifest as hackathons or social science research collaborations \cite{puthipiroj_structured_2025,universitat_hamburg_data-driven_nodate}. Students conduct empirical investigations, such as user testing, surveys and interviews, in order to align systems with the values of identified stakeholders and users.
This approach bridges theory and practice by evaluating case studies of concrete AI deployments. User experiences and values are incorporated when AI development receives and is informed by feedback from communities, stakeholders and users. Nevertheless, limitations include excessive focus on specific product development. This potentially constrains ethical evaluation horizons and creates unidirectional ethical judgment training. Furthermore, identifying and evaluating all relevant stakeholder experiences is complex. Moreover, this approach does not provide clear strategies for resolving conflicts between competing stakeholder values and needs, nor for addressing the underlying power dynamics inherent in technological contexts \citep{BleherBraun2023Reflections}.

\subsection{Challenges and critical gaps}

Together, these three approaches reveal both possibilities and inherent tensions of AI ethics education and in preparing students for real-world AI ethical challenges. Despite notable progress in integrating AI ethics into higher education, significant challenges persist from a conceptual-ethical perspective – challenges that extend beyond organizational or political considerations of implementation and skills training for AI literacy. Across these three observed approaches, certain limitations can be critiqued, from an ethical perspective: As with AI ethics approaches in general, first, AI ethics education programs, whether more or less top-down or bottom-up, tend to engage insufficiently with historical injustices, intersectional bias and the deep structures of power that shape technology and society. However, aligning technology with human rights requires that the experiences, vulnerabilities and perspectives of the most marginalized groups be not only reflected but integrated into AI ethics education to safeguard human dignity. 

Second, ethics education is often delivered through fragmented, stand-alone courses or certificate programs rather than being woven systematically throughout university and engineering curricula. While embedded ethics programs compellingly illustrate how the integration of engineering and ethical reflection can be operationalized through cross‑disciplinary collaboration and experiential learning \citep{laupichler_artificial_2022,lee_analyzing_2021,shih_learning_2021}, their scope remains inherently limited. Achieving fair and responsible AI design demands not only bridging disciplinary boundaries at the course level, but also embedding AI ethics education in multi‑stakeholder, multi‑domain and real‑world contexts. Preparing students for responsible AI engineering and citizenship that safeguard human rights requires structural engagement with actors across politics, industry, civil society and diverse application domains, ensuring that ethical reflection is continuously informed by practical deployment realities.

Third, institutional support for AI ethics education is inconsistent and often dependent on political circumstances, as demonstrated by recent events in the United States. Limited political and commercial commitment can undermine initiatives, and persistent resource constraints and insufficient accountability mechanisms can exacerbate these difficulties. Often, the continuity of ethics programs in higher education contexts depends on unstable political contexts, short-term funding or the voluntary involvement of individuals rather than sustained strategic institutional investment. Current political and international guidelines, however, do not adequately address the financial, organizational and human resources needed to establish AI ethics education sustainably standardized, effective, and implement it independently within academic institutions.

\subsection{Toward transformative integration}

Building on the challenges outlined above, the next step for higher education is to move beyond siloed or single‑method programmes towards a transformative integration of AI ethics education. Furthermore, a central consideration for developing AI ethics governance in higher education, as proposed in this paper, is the adoption of a broad societal perspective on AI education. In this context, considerations of citizenship and the common good must be placed at the forefront of AI ethics governance. 

This entails, first, preparing students for responsible AI engineering by equipping them for active participation in social life, civic engagement and democratic processes. Effective AI ethics education governance must therefore be oriented not only towards fostering technical and ethical competence, but also towards cultivating the values and skills necessary to shape AI in the service of society. This requires universities to foreground competencies and perspectives that critically address, for example, issues of power and inequality, in order to recognize and engage with the realities of marginalized communities, support those in disadvantaged positions, and contribute to overcoming injustices as well as preventing the misuse of AI technologies.

In light of this, second, establishing partnerships with community stakeholders should form a core part of the design and evaluation of AI ethics education curricula, rather than being considered supplementary additions. For example, concepts like service learning in non-profit or welfare organizations could foster social and ethical reflective competencies in students (Salam et al., 2019).

Third, to provide broader perspective, integrated governance frameworks for AI ethics education should establish systematic links between classroom learning, institutional decision‑making, policy development and accountability mechanisms. These connections may be realized, for example, through the use of policy‑making simulation models or the creation of real‑world AI deployment laboratories. 

The aim of such an approach extends beyond the design of interdisciplinary curricula: it incorporates the applied reflection of AI ethics within societal contexts, thereby generating empirically grounded insights into the real‑world consequences, opportunities and vulnerabilities associated with AI systems. In this trajectory, ongoing community engagement must be prioritized, ensuring that ethical education is not only well-intentioned but effective and enduring. Only by uniting technical expertise with socially reflective perspectives on justice, equity and democracy and by implementing practice-oriented, participatory and critically engaged methods, can higher education effectively foster and challenge ethical attitudes in AI \citep{lee_analyzing_2021,wiese2025ai}. Such an approach empowers students to shape the future of AI as responsible and engaged citizens. This enables them to shape AI's future through responsible citizenship.

\section{Conclusion}

This paper has argued that AI governance education in higher education must move decisively beyond siloed, principle-only, or tool-centric approaches. We mapped the limitations of the “BAG” separation (Build-Assess-Govern), showed how operational failures stem from mismatches between intention and outcome across five recurring domains (representation, specification, generalization, interaction, governance) and synthesized emerging best practices that integrate technical literacy, ethics, law and policy. The result is a modular, interdisciplinary curriculum that treats governance not as an add-on but as the organizing logic of learning, teaching students to anticipate failure modes, weigh trade-offs and translate abstract values into institutional and technical decisions.

The central claim is straightforward: effective AI governance education must be integrative, participatory and practice-oriented. Integrative means ethics, law and policy are threaded through the technical stack; from data collection and model design to deployment and monitoring, rather than cordoned off in separate courses. Participatory means students, communities, industry and policymakers co-produce learning goals and artifacts, ensuring that curricula do not drift into ethics-washing or ignore marginalized perspectives. Practice-oriented means repeated exposure to real or realistic cases, where students must make and justify choices under uncertainty, document compliance, evaluate externalities and propose adaptive mitigations.

Our comparative review of regulatory regimes underscores why this integration matters. The EU’s risk-based framework, China’s evolving data and platform rules and sectoral approaches elsewhere present moving, sometimes conflicting targets. Graduates will need fluency across jurisdictions, comfort with ambiguity and habits of continuous update. Embedding casework on compliance, documentation and accountability prepares students for this reality while sharpening their ethical reasoning and institutional literacy.

The proposed semester plan and module architecture in Appendix~\ref{app:course-structure} operationalize these insights. Technical foundations anchor understanding of model behavior, operational-failure taxonomy builds diagnosis and foresight, law and policy modules cultivate navigational competence and embedded-ethics and stakeholder-engagement activities foster deliberation skills. The capstone’s experiential focus; combining stakeholder consultation, risk assessment, regulatory documentation and policy simulation; turns knowledge into capability.

Still, we recognize limits. Curricula alone cannot guarantee responsible AI; incentives, governance capacity and institutional commitments matter. Faculty development remains a bottleneck and sustained partnerships with communities and regulators require resources and trust. To mitigate these gaps, we recommend universities commit to recurring instructor training, create credit-bearing community partnerships and establish governance sandboxes or “deployment labs” that connect teaching with institutional decision-making and external oversight.

The contribution of this paper is twofold: first, a synthesis of literature that situates AI ethics within a broader  AI governance perspective, centered on operational reality; second, a concrete, adaptable course design that any institution can implement and iterate upon. By aligning pedagogy with the life cycle of AI systems and the dynamics of real-world deployment, the curriculum cultivates professionals who can reason across disciplines, engage stakeholders, navigate regulation and design for justice, safety and accountability.

Looking ahead, three priorities stand out. First, build evidence: embed rigorous assessment of learning outcomes (ethical reasoning, governance documentation quality, stakeholder engagement efficacy) and publish results to refine the field. Second, connect levels of governance: link classroom artifacts (risk registers, model cards, DPIAs, audit plans) to institutional policies and sector standards to shorten the path from learning to practice. Third, broaden participation: co-create modules with civil society, public agencies and underrepresented communities so that governance education remains grounded, legitimate and responsive.

When universities embrace this integrated model, they can graduate a generation capable of steering AI toward public value; rather than merely coping with its externalities. The measure of success will not be perfect systems, but better institutions: ones that recognize and manage misalignment, surface and reduce power asymmetries and adapt as technologies and norms evolve. Education is the leverage point. Done well, it can turn AI governance from a compliance exercise into a civic and professional practice; one that is technically informed, ethically grounded and accountable to the societies AI is meant to serve.

\newpage
\bibliography{library}

\newpage
\appendix
\section{Exemplary Course Structure}
\label{app:course-structure}
The curriculum combines technical, ethical, legal and policy content in an
integrated, non-siloed manner. It addresses operational, societal and regulatory
challenges and prepares students for real-world, interdisciplinary work.

The following themes should be addressed throughout the modules:
\begin{itemize}
  \item \textbf{Justice, equity and inclusion:} Addressed explicitly in every module; critical engagement with power and bias.
  \item \textbf{Community engagement:} Partner with NGOs, industry, marginalized communities.
  \item \textbf{Faculty development:} Regular interdisciplinary training for instructors.
  \item \textbf{Assessment:} Combination of formative, summative and stakeholder-based feedback.
\end{itemize}

\subsection*{Example Semester Plan}

\begin{table}[h!]
\begin{center}
\renewcommand{\arraystretch}{1.3}
\begin{tabular}{p{2cm} p{4cm} p{7cm}}
\toprule
\textbf{Week} & \textbf{Module} & \textbf{Activities} \\
\midrule
1--2  & Introduction to AI \& society & Case discussions, foundational concepts \\
3--5  & Technical foundations & Labs, explainability exercises, technical demos \\
6--8  & Operational challenges & Bias audits, scenario analysis \\
9--10 & Law \& regulation & Moot court, global comparison, regulatory briefings \\
11--12 & AI ethics foundations & Principle debates, justice analysis \\
13--14 & Interdisciplinary approaches & VSD workshop, stakeholder interviews \\
15--16 & Experiential project & Team project, policy simulation, stakeholder feedback \\
17    & Reflection \& adaptive governance & Capstone presentations, self and peer assessment \\
\bottomrule
\end{tabular}
\end{center}
\end{table}

\subsection*{Modules}

\paragraph{Module 1: Introduction to AI \& society}
\begin{itemize}
  \item History and paradigms of AI (symbolic, probabilistic, machine learning, deep learning, etc.)
  \item Societal impact of AI: opportunities, risks and urgency for ethical governance
  \item Interdisciplinary nature of AI ethics
  \item Key case studies (e.g., Amazon recruitment AI, COMPAS, LLMs in society)
\end{itemize}
\emph{Learning outcomes:}
\begin{itemize}
  \item Foundational AI literacy
  \item Critical reflection on AI’s social impact
\end{itemize}

\paragraph{Module 2: Technical foundations of AI}
\begin{itemize}
  \item Types of AI: symbolic, probabilistic, machine learning (supervised, unsupervised, reinforcement), evolutionary, hybrid
  \item Architectures: neural networks, CNNs, RNNs, transformers, RAG, CAG, multi-agent and hybrid systems
  \item AI development lifecycle: from problem definition and data collection to deployment and monitoring
  \item Concepts: explainability, bias, robustness, adversarial attacks, transparency
\end{itemize}
\emph{Learning outcomes:}
\begin{itemize}
  \item Analyze technical aspects and ethical implications
  \item Understand the lifecycle of AI development
\end{itemize}

\paragraph{Module 3: Operational challenges \& governance failures}
\begin{itemize}
  \item Taxonomy of operational failures:
    \begin{itemize}
      \item Representation (bias, feedback loops)
      \item Specification (proxy goals, confounders)
      \item Generalization (Goodhart’s Law, OOD data)
      \item Interaction (user overtrust, misuse)
      \item Governance (oversight, adversarial risks)
    \end{itemize}
  \item Real-world case studies: predictive policing, AI in healthcare, LLMs, etc.
\end{itemize}
\emph{Learning outcomes:}
\begin{itemize}
  \item Diagnose operational failure
  \item Understand the need for governance
\end{itemize}

\paragraph{Module 4: Law, regulation and international approaches}
\begin{itemize}
  \item Comparative regulation: EU AI Act, Digital Services Act, Digital Markets Act, GDPR, China’s PIPL and data laws, US and other frameworks
  \item Risk-based approaches and regulatory categorization
  \item Extraterritoriality and global effects
  \item Data rights, intellectual property and copyright
\end{itemize}
\emph{Learning outcomes:}
\begin{itemize}
  \item Navigate and compare regulatory frameworks
  \item Anticipate compliance and ethical issues
\end{itemize}

\paragraph{Module 5: Foundations of AI ethics}
\begin{itemize}
  \item Major ethical principles and frameworks (EU HLEG, UNESCO, IEEE, etc.)
  \item Core topics: fairness, accountability, transparency, privacy, robustness, safety, human agency
  \item Approaches: top-down (principle-based), bottom-up (contextual), participatory (stakeholder engagement)
  \item Critical perspectives: ethics washing, power structures, justice oriented analysis
\end{itemize}
\emph{Learning outcomes:}
\begin{itemize}
  \item Apply and critique ethical frameworks
  \item Recognize limitations and gaps in practice
\end{itemize}

\paragraph{Module 6: Interdisciplinary \& participatory approaches}
\begin{itemize}
  \item Value Sensitive Design (VSD): stakeholder mapping, participatory assessment
  \item Embedded Ethics: integrating ethics in technical projects/courses
  \item Principle-Aligned Ethics: building on institutional and international standards
  \item Strengths and limitations of each approach; case examples from leading universities
  \item Community partnerships and stakeholder engagement exercises
\end{itemize}
\emph{Learning outcomes:}
\begin{itemize}
  \item Design and participate in interdisciplinary ethical deliberation
\end{itemize}

\paragraph{Module 7: Experiential \& Project based learning}
\begin{itemize}
  \item Team projects with real or simulated AI deployments:
    \begin{itemize}
      \item Stakeholder analysis and consultation
      \item Technical implementation with ethical risk assessment
      \item Regulatory compliance documentation
      \item Policy proposal or white paper
    \end{itemize}
  \item Role-playing exercises (regulatory hearings, community forums, company boards)
\end{itemize}
\emph{Learning outcomes:}
\begin{itemize}
  \item Practical skills in real-world AI governance
  \item Teamwork across disciplines
\end{itemize}

\paragraph{Module 8: Reflection, evaluation and adaptive governance}
\begin{itemize}
  \item Curriculum feedback and iterative improvement
  \item Peer/self-assessment in ethical reasoning and governance skills
  \item Capstone reflection connecting theory, practice and professional values
\end{itemize}
\emph{Learning outcomes:}
\begin{itemize}
  \item Build reflexivity and adaptive learning
  \item Prepare for continuous change in the field
\end{itemize}

\subsection*{Course Outcomes}
\begin{itemize}
  \item Identify, analyze and address ethical, legal and operational risks in AI systems.
  \item Work across disciplines and with stakeholders for responsible AI solutions.
  \item Navigate and critique major international regulatory frameworks.
  \item Reflect on professional responsibilities and AI’s broader societal impacts.
  \item Develop practical and adaptive skills for leadership in AI governance.
\end{itemize}

\subsection*{Implementation Notes}
\begin{itemize}
  \item \textbf{Integration:} Ethics, law and governance are woven throughout technical content.
  \item \textbf{Project based:} Experiential, “learning by doing” focus.
  \item \textbf{Instructor diversity:} Team-teaching across disciplines.
  \item \textbf{Continuous update:} Content evolves with new research and policy.
\end{itemize}

\end{document}